\documentclass[epj]{svjour}
\usepackage{graphics}
\usepackage{amsmath,amssymb}
\usepackage{cuted}
\usepackage{verbatim}

\usepackage{graphicx}
\usepackage{xcolor}
\usepackage{lipsum}
\usepackage{comment}

\begin{document}
\title{On the emergence of a classical Isotropic Universe from a Quantum $f(R)$ Bianchi Cosmology in the Jordan Frame}

\author{Mariaveronica De Angelis\inst{1}\thanks{ mdeangelis1@sheffield.ac.uk} \and Giovanni Montani\inst{2}\inst{3}
\thanks{giovanni.montani@enea.it}%
}                     
\institute{School of Mathematics and Statistics (SoMaS), University of Sheffield, Hounsfield Road, Sheffield, S3 7RH, United Kingdom \and ENEA, Fusion and Nuclear Safety Department, C.R. Frascati, Via E. Fermi 45 (00044) Frascati (RM), Italy \and Physics Department (VEF), Sapienza University of Rome, P.le A. Moro 5 (00185) Roma, Italy }
\date{Received: date / Revised version: date}
%
\abstract{
We demonstrate a spontaneous tendency of quantum wave packets to become quasi-classical, providing a classical limit for the Universe dynamics. However, this limit is not maintained in
the future (after a critical value of the relational time) and a spreading process is turned on. We show that the onset of an inflationary scenario is not able to make this localization stable of the wave packets for the Bianchi I model. Instead, when we implement a perturbative inflationary scenario for the isotropic Universe a mechanism of
stable classicalization of the Universe emerges.
This result outlines a sharp difference between the standard relativistic cosmology and a modified $f(R)$ paradigm.
\PACS{
      {PACS-key}{discribing text of that key}   \and
      {PACS-key}{discribing text of that key}
     } 
} 
\maketitle
\section{Introduction}

It is well-known \cite{BKL82,montani95,primordial} 
that the implementation of General Relativity 
to the cosmological problem leads to
the emergence of a  primordial singularity which is commonly dubbed Big-Bang 
\cite{winberg2008} for the isotropic Universe. This fact suggests that
for a sufficiently high value of the space-time curvature and matter-energy density, the predictability of the Einstein equation becomes questionable. 
To reformulate the dynamics of the gravitational field in a singularity-free picture, two main approaches have been pursued over the years. On one hand, it was argued that the Einstein-Hilbert action must be replaced by a more general setting, with particular reference to the simplest one 
of a metric $f(R)$ gravity \cite{Odintsov,Sot}; on the other 
hand, it was suggested that a quantum dynamics of the gravitational field must replace the classical Einsteinian 
picture \cite{DeWitt1,d2,d3}, especially given a canonical quantization procedure of the metric field \cite {Kuchar81}.

The present analysis investigates the 
physical implication of joining together these two points of view, \emph{i.e.} our theoretical framework to study the dynamics of the Bianchi Universes 
\cite{MontaniBattBenImp,BKL70,BenMont07} 
namely, we canonically quantize a metric f(R) theory 
as viewed in the so-called \emph{Jordan frame} \cite{Odintsov,Sot,Oik}. 

This both classical and quantum reformulation of the Einsteinian picture has also a specific motivation in the problem of time, which affects the canonical method of quantizing the gravitational field \cite{Isham92}. Indeed, the Hamiltonian function of the gravitational field identically vanishes \cite{adm} so that the resulting quantum dynamics, described in terms of the Universe wave functional, 
appears as frozen. 
The most widely accepted solution to this problem consists of the so-called \lq \lq relational time" approach \cite{Isham92,rovelli}, \emph{i.e.} a physical clock is constructed via an internal gravity or matter degree of freedom \cite{CanQuantumGrav}. Some examples of how to construct a time variable for the quantum gravity dynamics via a specific field are \cite{CianfMontZon,CianfBombMont,ManicciaDea,thiemann,Kiefer}. 

When the canonical method of 
quantization is applied to the metric $f(R)$ theory in the \emph{Jordan frame}, the physical clock is offered by the non-minimally coupled scalar field, 
which is naturally present in such a formulation. 

This point of view was first explored in \cite{Costantini}, where the cosmological problem has been formulated for the $f(R)$ gravity in the \emph{Jordan frame} with particular reference to 
the Bianchi Universes. There, the question concerning the canonical quantum evolution of the primordial Universe in terms of the non-minimally coupled scalar field has been also discussed in some detail. 

Furthermore, in \cite{DeaFigMont}, it was 
analysed the reduced phase space quantum dynamics for an isotropic Universe in the same conceptual paradigm by showing that, if the non-minimally coupled scalar field 
plays the role of a clock a very peculiar feature emerges: a wave packet, evolved by the corresponding Schr\"odinger equation, manifests a progressive localization as the Universe expands. This surprising feature is however not stable along the Universe dynamics, since a delocalization phenomenon, soon or later, 
will star again. There, it was 
argued that a possible mechanism to make 
stable such a process of Universe 
classicalization could emerge during a 
de Sitter phase, typical of an inflationary scenario \cite{Kolb-Turner}.

The present study continues and completes 
the investigation pursued in \cite{DeaFigMont} by generalizing the dynamics to the anisotropic homogeneous Bianchi Universes and by demonstrating that the conjecture that the de Sitter phase can stabilize the Universe classicalization is well-grounded.

We first analyse under which conditions the dynamics of a Bianchi model, as viewed in a $f(R)$ theory in the \emph{Jordan frame}, is reducible essentially to the dynamics of a Bianchi I model in the presence of a non-minimally coupled scalar field. In other words, we argue that in agreement with the analysis in \cite{Costantini}, both the Universe spatial curvature (responsible for a potential term in the Bianchi Universe Hamiltonian) and the potential term of the scalar mode (fixed by the specific form of the $f(R)$ function) can be neglected up to the first approximation. 

Based on this statement, we study the quantum dynamics of the Bianchi I model both in a vacuum and in the presence of the on-set of an inflationary paradigm 
by clarifying that the classicalization process (the standard deviation of the density probability significantly decreases) is always present with similar features to those one discussed for the isotropic Universe in \cite{DeaFigMont}. 
However, in both the cases mentioned above, the localized wave packet is, soon or later, subjected to subsequent spreading dynamics, \emph{i.e.} the Universe becomes 
a fully quantum system. 

The analysis is then completed by demonstrating that the presence of a cosmological constant in the quantum dynamics (mimicking an inflationary phase) of the isotropic Universe is really the mechanism responsible for the emergence of a 
stable classical one from the Planckian era, 
\emph{i.e.} no subsequent spreading of the wave packets after the localization takes place. 

This result leads us to infer that also 
for a Bianchi I Universe, the presence of 
cosmological dynamics could imply the 
same stability of the emergent classical dynamics. 
In particular, we argue that such a classicalization of the anisotropic Bianchi I model passes before through quasi-isotropization dynamics and then 
the classical isotropic limit is approached. More specifically, as discussed for standard gravity in \cite{Muccino}, the presence of a cosmological constant term in the quantum dynamics is expected to induce an isotropization process 
on average ( the anisotropic variables are 
essentially suppressed). From this stage, the classical limit is obtained similarly to what we show here for the exact isotropic Universe.

The manuscript is structured as follows. In Sec.(\ref{sec2}), we present the modified theory of gravity $f(R)$ in scalar-tensor representation, focusing the attention on the \emph{Jordan frame} in which the non-minimally coupled scalar field to gravity is present. In Sec.(\ref{secIII}), we consider the $f(R)$ theories in scalar-tensor representation to achieve the Hamiltonian formalism of gravity for homogeneous and anisotropic Universes, namely the Bianchi models. In Sec.(\ref{secIV}), we analyse the behaviour of a Bianchi I cosmology in the paradigm of Schr\"odinger-like formulation of canonical quantum gravity. Moreover in Sec.(\ref{secV}), we investigate whether introducing the cosmological constant, the analysis of the Bianchi I model in the vacuum can be modified. Furthermore, in Sec.(\ref{FRLWsec}) we repeat the analysis of the quantum dynamics for the FRLW model in the presence of a perturbative cosmological constant to clarify how the isotropic model is subjected to a stable process of localization, approaching a classical expanding Universe. Finally, in sec.(\ref{secVII}) conclusions are drawn.

\section{F(R) theories in the Jordan frame}\label{sec2}

Einstein's General Relativity is largely accepted as the fundamental theory for describing the geometrical properties of space time. Actually, the Einstein-Hilbert action is only the most simple proposal (providing second-order field equations, see the Lovelock theorem \cite{Lovelock}) and even if the determination of the gravitational field kinematics (tensor formalism) is a naturally consistent formulation, different schemes are admitted by the dynamics of the gravitational field.
One of the simplest modifications to GR is the $f(R)$ gravity in which the Lagrangian density $f$ is an arbitrary function of $R$ \cite{Odintsov,Ruzmaikina,Bergmann,Buchdahl,Capozziello}. In order to derive Einstein's field equations, in literature two main approaches are present: affine and metric formulation \cite{Sot,Gonzalo} and in this paper, we will take into consideration the last one.
We start with the 4-dimensional action in $f(R)$ gravity 
\begin{equation}
S=-\frac{1}{2\kappa}\int d^4x \sqrt{-g}\ f(R) + \int d^4x \,\mathcal{L}_M (g_{\mu\nu}, \Psi_M),
\label{action}
\end{equation}
where $\kappa \equiv 8 \pi G$ ($G$ being the Newton constant, using c=1), $g$ is the determinant of the metric tensor $g_{\mu \nu}$ and $\mathcal{L}_M$ is a matter Lagrangian which depends on the matter fields $\Psi_M$ and the metric too. The field equations can be derived by varying the action (\ref{action}) with respect to $g_{\mu \nu}$ and they are characterized by fourth-order differentiation
\begin{equation}
f(R)'R_{\mu \nu}-\frac{1}{2}f(R)g_{\mu \nu}-[\nabla_{\mu}\nabla_{\nu}-g_{\mu \nu}\Box]f'(R)=\kappa T_{\mu \nu},
\end{equation}
where $T_{\mu \nu}$ is the matter source energy momentum tensor, $f'(R)\equiv df/dR$, $\nabla_{\mu}$ the covariant differential derivative with respect to the metric $g_{\mu \nu}$ and $\Box \equiv g^{\mu \nu}\nabla_{\mu}\nabla_{\nu}$ is the d'Alambert operator in curved manifolds. 
We can introduce a new field $\chi$ and write the dynamically equivalent action \cite{Teyssandier}
\begin{equation}
S=-\frac{1}{2\kappa}\int d^4x \sqrt{-g}(f(\chi)+f'(\chi)(R-\chi))+ S_M(g_{\mu\nu}, \Psi_M).
\end{equation}
Variation with respect to $\chi$ leads to the equation $\chi=R$ if $f''(\chi)\neq 0$ \footnote{this condition can be relaxed by requiring the injectivity of the function $f$, see \cite{olmo}.}, which reproduces action (\ref{action}). Now, by redefinition of the field $\chi$ into $\xi=f'(\chi)$ and collecting

\begin{equation}
V(\xi)=\chi(\xi)\xi-f(\chi(\xi)),
\end{equation}
the action reads as
\begin{equation}
S_J=-\frac{1}{2\kappa}\int d^4x \sqrt{-g}(\xi R-V(\xi))+S_M(g_{\mu \nu},\Psi_M),
\end{equation}
and it is known as the $f(R)$ gravitational action in the \textit{Jordan frame}. Thus, we get a dynamical scalar field which is non-minimally coupled with the curvature $R$. Hence, field equations turn out to be
\begin{equation}
\begin{cases}
G_{\mu \nu}=\frac{\kappa}{\xi}T_{\mu \nu}-\frac{1}{2 \xi}g_{\mu \nu}V(\xi)+\frac{1}{\xi}(\nabla_{\mu}\nabla_{\nu}\xi-g_{\mu \nu}\Box \xi) \\R=V'(\xi)
\label{secondorder}
\end{cases}
\end{equation} 
corresponding now to a second-order formulation. Taking the trace of the first equation (\ref{secondorder}) and using the second one
\begin{equation}
3\Box \xi +2V(\xi)-\xi\frac{dV}{d\xi}=\kappa T_{\mu}^ {\mu},
\end{equation}
that determines the dynamics of the scalar field for a given source of matter.

Above, we described the general structure of the $f(R)$ model in the \emph{Jordan frame} also in the presence of matter but it is
worth stressing that, since from the very
beginning, the following analysis is performed always neglecting both the thermal bath (radiation) energy density as well as the potential term of $\xi$ which is fixed by the specific form of the function $f(R)$, for a justification in
the context of the Bianchi Universe,
see the end of the next section.

\section{Bianchi Universes in the Jordan frame}\label{secIII}

In this section, we analyse the Hamiltonian formulation of the 
homogeneous Bianchi Universes \cite{primordial,MontaniBattBenImp,Misner}, in order to provide a general theoretical 
framework interpreting the quantum dynamics of the Bianchi I model developed below.

In the diagonal formulation of a Bianchi cosmology, we deal with the following line element \cite{Gravitation}
\begin{equation}
ds^2 = N^2dt^2 - e^{\alpha}
\left( e^{\beta}\right)_{ab}
\omega^a\omega^b
\, ,
\label{demo1}
\end{equation}
where $\beta \equiv diag \{ 
\beta_+ + \sqrt{3}\beta_-, \beta_+ - 
\sqrt{3}\beta_-, -2\beta_+\}$ is a traceless matrix and $\omega^a$ ($a,b=1,2,3$) denote the 1-forms characterizing the isometry group of one of the nine Bianchi models \cite{LL}.
Due to the homogeneity constraint, 
the lapse function $N$ and the Misner 
variables $\alpha,\beta_{\pm}$ 
\cite{Gravitation} are all functions of time only and, it is rather straightforward to check that in the \emph{Jordan frame} of a 
$f(R)$ modified gravity and in the presence of a cosmological constant $\Lambda$ \cite{Costantini,DeaFigMont}, the 
Bianchi Universes are associated to the 
following reduced ADM-action \cite{primordial,Gravitation}
\begin{equation}
S_B = \int d\xi \left( 
p_{\alpha}\frac{d\alpha}{d\xi} + p_+\frac{d\beta_+}{d\xi} + 
p_-\frac{d\beta_-}{d\xi} - 
H_{ADM}\right)
\, , 
\label{demo2}
\end{equation}
where we adopted the non-minimally 
coupled scalar field as a time variable. The associated reduced Hamiltonian reads as
\begin{strip}
	\begin{equation}
	H_{ADM} \equiv \frac{1}{\xi}\left( p_{\alpha} + \sqrt{p_{\alpha}^2 - p_+^2 - p_-^2 - 6\xi e^{2\alpha}V_B(\beta_{\pm}) - 6\xi e^{3\alpha}\left( V(\xi ) + 
		\Lambda\right)}\right)
	\, , 
	\label{demo3}
	\end{equation}
\end{strip}
with $p_{\alpha}$ and $p_{\pm}$ 
denoting the conjugate momenta to the 
corresponding variables and $V_B$ 
is fixed by the isometry of the considered Bianchi model (we adopt geometrical units and have been integrated, without loss of generality, on a unit space volume). 
It is worth stressing that, the ADM-reduction procedure requires the following specification of the temporal gauge
\begin{equation}
N = N_{ADM}\equiv \frac{3e^{3\alpha}}{2\left(\xi H_{ADM} - p_{\alpha}\right)}
\, . 
\label{demo4}
\end{equation}
Now, it is well-known \cite{Kirillov97,ImponenteMontani2001} that, 
for the Bianchi IX model (the most general Bianchi Universe with an isotropic limit), near the initial singularity, \textit{i.e. $\alpha \rightarrow -\infty$}, the potential term admits the representation
\begin{equation}
e^{2\alpha}V_B \propto D^{2H_1} + D^{2H_2} + D^{2H_3},
\label{demo5}
\end{equation}
where
\begin{align*}
D &= e^{3\alpha},\\
H_1 &= \frac{1}{3}+\frac{\beta_+ - \sqrt{3}\beta_-}{3\alpha},\\ 
H_2 &= \frac{1}{3}+\frac{\beta_+ + \sqrt{3}\beta_-}{3\alpha},\\
H_3 &= \frac{1}{3}-\frac{2\beta_+}{3\alpha}. 
\end{align*}
Thus, asymptotically to the singularity $D\rightarrow 0$, this term leads to a triangular potential well, in which walls are defined by the simultaneous conditions $H_a\equiv0$ ($a=1,2,3$) and receding with the decreasing of $\alpha$ towards $-\infty$. 

\begin{figure}[h!]
	\centering
	\includegraphics[width=9cm]{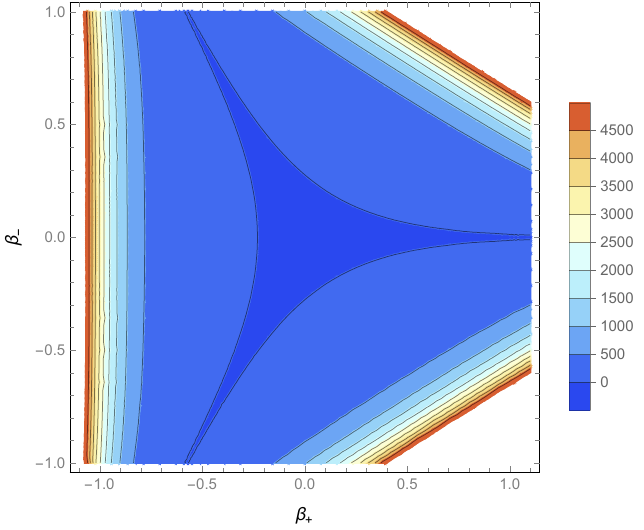} 
	\caption{Equipotential lines for the Bianchi IX potential in the plane ($\beta_+$, $\beta_-$).}
	\label{triangolo}
\end{figure}

While in General Relativity the Bianchi IX Universe is chaotic, in \cite{Costantini} it has been shown that for $f(R)$ theories considering $V(\xi)\sim \xi^n$, with $n<6$, the chaos of the asymptotic dynamics is no longer present and the so-called \lq \lq Kasner Stability Region" \cite{BK73} emerges. By other words, in the mentioned theories, the singularity (for a bouncing Bianchi I cosmology and other quantum formulations of the Bianchi Models see \cite{flavio,GabEleonoraI,Ashtekar,giovamontschiatt,giovmont2,battistimont}) is reached by a Kasner solution de facto corresponding to a Bianchi I (potential-free) cosmology. 
More in general, a Kasner-like evolution takes place each time the potential walls (associated to the spatial curvature of the Universe) are negligible, like in the centre of the triangular configuration Fig(\ref{triangolo}).

In this respect, within the framework of classical General Relativity and in the Wheeler-DeWitt scenario, in \cite{Muccino} and \cite{Kirillow2002} respectively, it has been investigated the situation in which the cosmological constant term dominates the potential one, mimicking the trigger of an inflationary dynamics. 

Here, we analyse the behaviour of a Bianchi I cosmology without and with the presence of the cosmological constant in the paradigm of Schr\"odinger-like formulation of canonical quantum gravity.
Thus, the analysis presented in the following sections must be intended as a specific limit of the Bianchi IX Universe evolution in which the momentum contribution (the kinetic component) of the Hamiltonian (\ref{demo3}) dominates the potential terms. 
In particular, we are always neglecting the contribution due to $V(\xi)$ taking into account its suppression near the singularity by the small value of the factor $e^{3\alpha}$. And, in the most interesting case when a cosmological term is present, our Bianchi I study must be thought of as in correspondence to the validity of the following inequalities
\begin{equation}
\Lambda \gg V(\xi),\ \ \  
\Lambda \gg e^{-\alpha}V_B. 
\label{demo6}
\end{equation}

We stress that the second of these conditions is favoured by the increasing value of $\alpha$ as the Universe expands, which we will see below that corresponds to the decreasing values of $\xi$ towards zero in the case of a Bianchi I evolution.
However, this consideration relies on the idea that the behaviour of the quantum anisotropies is under control, \emph{i.e.} the average value of the potential term $V_B$ is stable and it is also a slow function of time.

The possibility to neglect the potential term $V(\xi )$ in the classical and
quantum dynamics is rather natural towards the initial singularity ($\alpha \rightarrow - \infty$), since its contribution is suppressed by the
exponential decrease in the volume.
Thus, if not valid for any $f(R)$ functional form (see the classical analysis of the Bianchi IX model
in \cite{Costantini}),
the idea of dealing, near enough to
the singularity, with a potential-free non-minimally coupled scalar field, appears well-grounded and in the quantum sector, it should be regarded as a viable assumption.
Even more well-posed is the negligibility of the thermal bath (radiation) energy density concerning the anisotropy Hamiltonian kinetic term since the former diverges near the singularity as the inverse fourth power of the cosmic scale factor, while the latter diverges as the inverse sixth power and therefore it will dominate the dynamics.
The situation is quite different when we consider an inflationary paradigm which is associated with a rapid expansion of the Universe due to the cosmological constant term, \emph{i.e.} the vacuum energy.
Indeed, at least on a classical level, the possibility to neglect the potential term of $\xi$ has to break down as a result of a large Universe volume. It is just for this reason that, below, the Bianchi I and the isotropic Universe
dynamics in the presence of a cosmological constant are analysed on a perturbative scheme only,
\emph{i.e.} before that the cosmological constant
starts to really dominate the system evolution.
Also, the quantum analysis of the isotropic inflationary  model must be intended as the quantum on-set of a de Sitter phase which can
generate a classical configuration.
It is clearly expected that after
the Universe classicalization, the subsequent inflationary behaviour
has to be, soon or later, influenced
by the specific form of the considered
$f(R)$ model, but this part of the
evolution is out of the purposes of this study.
Instead, the possibility to neglect
the thermal bath contribution is
still a natural assumption of the
inflationary paradigm, since the
dynamics is by hypothesis dominated
by the vacuum energy of the Universe
phase transition \cite{primordial,Kolb-Turner}.

\section{Schr\"odinger dynamics of Bianchi I model}\label{secIV}

Quantum Cosmology aims to provide a quantum description for homogeneous cosmological models. However, a major issue to struggle with is the problem of the time. It relies on the identification of a proper time-like variable avoiding the \emph{frozen formalism} which leads to different results whether it is tackled before or after quantization \cite{guven-1992}. Two approaches have been pursued to give a meaningful interpretation of the wave function of the Universe in a probabilistic way. In the first one, the \emph{super-Hamiltonian} constraint is classically solved and the resulting Schr\"odinger equation is quantized. This is the so-called reduced phase space quantisation (RPSQ) that is the most straightforward procedure since it is an exact algorithm requiring no WKB approximation based on the wave function of the Universe, even if the Hamiltonian density can be non-local. The second one is achieved by implementing both the WKB and Born-Oppenhaimer approximations, namely, the Vilenkin approach, for details see \cite{ManicciaDea,DeaMont}.

In the following analysis, we consider the RPSQ approach to recover the model dynamics, namely, we chose here the non-minimally coupled scalar field as a time variable before quantizing the system \cite{Isham92}.

After having solved the \emph{super-Hamiltonian} constraint without the cosmological constant term according to the ADM procedure, for the Bianchi I model we deal with the relation
\begin{equation}
p_{\xi}=\frac{1}{\xi}\biggl(p_{\alpha}+ \sqrt{p_{\alpha}^2-p_{+}^2-p_{-}^2} \biggr)=-h_{\xi},
\label{pxi}
\end{equation}
where $h_{\xi}$ coincides with $H_{ADM}$ in \eqref{demo3} once we neglect the potential contributions. As a direct consequence of implementing the classical Hamiltonian constraints on the operator level, we derive the equation \emph{a la} Schr\"odinger
\begin{equation}
i\frac{\partial \psi}{\partial \xi}=-\frac{1}{\xi}\biggl(-i\partial_{\alpha}+\sqrt{-\partial_{\alpha}^2+\partial_{+}^2+\partial_{-}^2}\biggr)\psi;
\end{equation}
the evolution of the so-called \lq \lq physical Hamiltonian", \emph{i.e.} $\hat{h_\xi}$, is therefore
described with respect to $\xi$ namely the variable canonically conjugate to $p_{\xi}$.

Now, making use of the plane wave 
\begin{equation}
\psi(\alpha,\beta_{\pm},\xi)=e^{i(k_{\alpha}\alpha +k_{+}\beta_+ +k_-\beta_-)}y(\xi)
\end{equation}

and by applying to it the square root derived from the non-local Hamiltonian \cite{Lammerzahl}, we obtain
\begin{equation}
i\xi \partial_{\xi}y(\xi)=\biggl(-k_{\alpha}-\sqrt{k_{\alpha}^2-k_+^2-k_-^2}\biggr)y(\xi),
\end{equation}
in which we denote the derivative for $_\xi$ by the corresponding subscript.
Hence, the solution of the equation for $y(\xi)$ is found to be
\begin{equation}
y(\xi)=\mathcal{B}\,\xi^{i(k_{\alpha}+\sqrt{k_{\alpha}^2-k_+^2-k_-^2})},
\end{equation}
with $\mathcal{B}$ being the integration constant (set to 1). The complete wave function can be written as
\begin{equation}
\psi(\alpha,\beta_{\pm},\xi)=e^{i(k_{\alpha}\alpha +k_{+}\beta_+ +k_-\beta_-)}\xi^{i(k_{\alpha}+\sqrt{k_{\alpha}^2-k_+^2-k_-^2})}.
\label{phi}
\end{equation}

The next step is to construct a localized wave packet by using a Gaussian distribution on $k_{\alpha}$, $k_{+}$ and $k_{-}$, \emph{i.e.}
\begin{equation}
\Psi(\alpha, \beta_{\pm}, \xi)= \int^{+ \infty}_{- \infty} dk_{\alpha}dk_{+}dk_{-} A(k_\alpha, k_{+}, k_{-})\psi(\alpha, \beta_{\pm}, \xi),
\end{equation}
with $A(k_{\alpha},k_+,k_-)$ being the product of three Gaussian functions in $k_{\alpha},k_+,k_-$ respectively
\begin{equation}
A(k_a)= \frac{1}{\sqrt{2\pi}\sigma_a}e^{-\frac{(k_a-\bar{k}_a)^2}{2\sigma_a^2}},
\end{equation}
where $\sigma_a$ ($a=\alpha, +,-$) is the standard deviation. Moreover, we denoted by $\bar{k}_a$ the mean value of the distribution.

Therefore, according to the natural scalar product for a Scr\"odinger-like equation, the normalizable probability density is
\begin{equation}
|\Psi|^2=\Psi \Psi^*,
\label{prob}
\end{equation}
which provides the probability of finding the universe at a certain instant $\xi$ for a unit of $\beta_+$, $\beta_-$ and $\alpha$. 

\begin{figure}
	\centering
	\includegraphics[width=8cm]{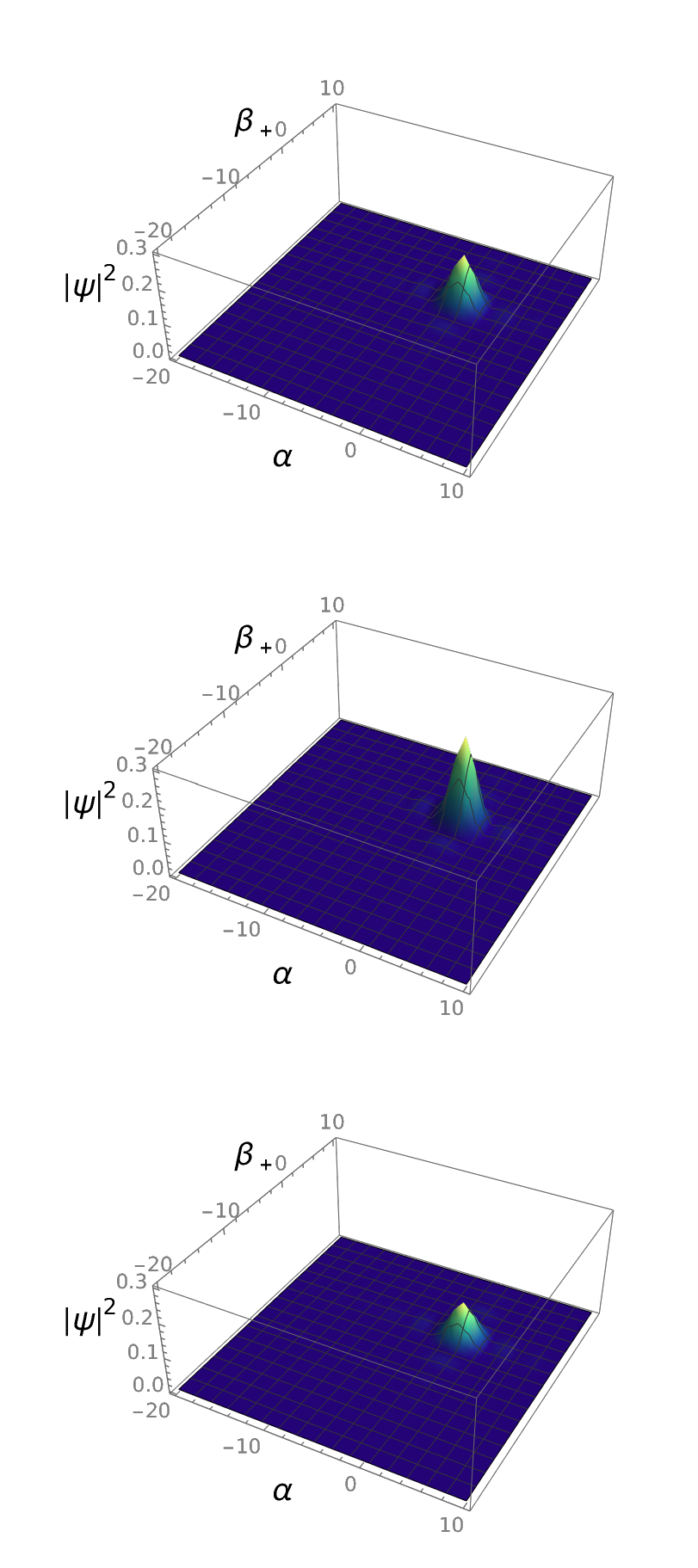} 
	\caption{Plots of $|\Psi|^2$ evaluated via numerical integration with working precision of 0.1. Starting from top, the probability density has been evaluated for different values of the relational time $\xi=10^3,10,1$ respectively. We adopted a suitable choose of $\beta_-= 5$.}
	\label{BianchiI}
\end{figure}

In Fig.(\ref{BianchiI}) we observe a very peculiar feature concerning a progressive localization of the wave packet from the singularity up to a precise value of the relational time $\xi$. To better clarify this behaviour let us consider the Universe volume related to $e^{3\alpha}$ for a fixed value of $\beta_-$ and different values of $\beta_+$. As it is shown, a behaviour of the wave packet approaching a localization namely a classicalization emerges. However, this classical state is not dynamically stable since between $\xi=10$ and $\xi=1$ the spreading mechanism starts leading to a de-localized wave packet, \emph{i.e.} a quantum picture again. It is worth noticing that it could suggest the introduction of the inflationary scenario as a possible candidate for maintaining a localized profile. Given this guess, the cosmological constant $\Lambda$ is considered in the section below.

Finally, we stress that our analysis concerns the very early phases of the Universe in which its dynamics are properly described by a canonical quantum picture, \emph{i.e.} all the physical information is contained in the Universe wave function. Thus, for this reason, the concept of space-time curvature is no longer directly applicable. Nonetheless, the space curvature and the extrinsic one of the Bianchi models can be still recovered as average quantities. Indeed, the former comes from the average value of the potential term present in the Hamiltonian (vanishing for the Bianchi
I model) and the latter can be reconstructed via the average values of the conjugate momenta to
the Misner variables. Clearly, these average quantities are expressed via the relational time, \emph{i.e.} the
scalar field emerging from the $f(R)$ theory in the \emph{Jordan frame} and their translation in terms of a
labelled time coordinate is not an immediate procedure when the classical limit has not still achieved.
However, the main issue of the following analysis is just the possibility to achieve a state for the Universe which is very close to the classical dynamics and, hence, the notion of classical space-time
curvature is naturally recovered in that limit.

\section{Quantum dynamics with a cosmological constant}\label{secV}
Here, we analyse the dynamics of the Bianchi I model considering the cosmological constant $\Lambda$. Hence, the relation (\ref{pxi}) becomes
\begin{equation}
p_{\xi}=\frac{1}{\xi}\Biggl( p_{\alpha}+\sqrt{p_{\alpha}^2-p_{+}^2-p_-^2-6\xi e^{3\alpha} \Lambda } \Biggl)= -h_{\xi}.
\label{pxiL}
\end{equation}

It is worth noting that, as in the previous
section and in agreement with the condition \eqref{demo6}, we are neglecting both the scalar curvature of a given Bianchi model, as
well as the potential term of the field
$\xi$. Actually, as discussed at the end of Sec.(\ref{secIII}) the former term can not be easily estimated in a full quantum regime, due to the evolution of the anisotropies $\beta_+$ and $\beta_-$. Thus, it can be reliably neglected only in the initial phase of
the inflationary evolution as considered below.
Therefore, here we describe the wavepackets dynamics
during the on-set of the inflation phase and,
also to make more stable the numerical
analysis, we expand the ADM-Hamiltonian
\eqref{pxiL} with respect to the
cosmological constant term, obtaining as Schr\"odinger equation
\begin{equation}
i\xi\partial_{\xi} y(\xi)=\left(-k_{\alpha}-K+ \frac{6\xi \Lambda e^{3\alpha}}{2K}\right)y(\xi),
\end{equation}
where $K\equiv\sqrt{k_{\alpha}^2-k_-^2-k_+^2}$ and its solution is
\begin{equation}
y(\xi)=\mathcal{C}e^{i\left(\frac{-3\xi\Lambda e^{3\alpha}+\log(\xi)(K^2+k_{\alpha}K)}{K}\right)},
\end{equation}
with $\mathcal{C}$ being the integration constant.

Hence, the complete wave function reads
\begin{equation}
\psi(\alpha,\beta_{\pm},\xi)=e^{i(k_{\alpha}\alpha+k_{+}\beta_++k_{-}\beta_-)}y(\xi).
\end{equation}

\begin{figure}
	\centering
	\includegraphics[width=8cm]{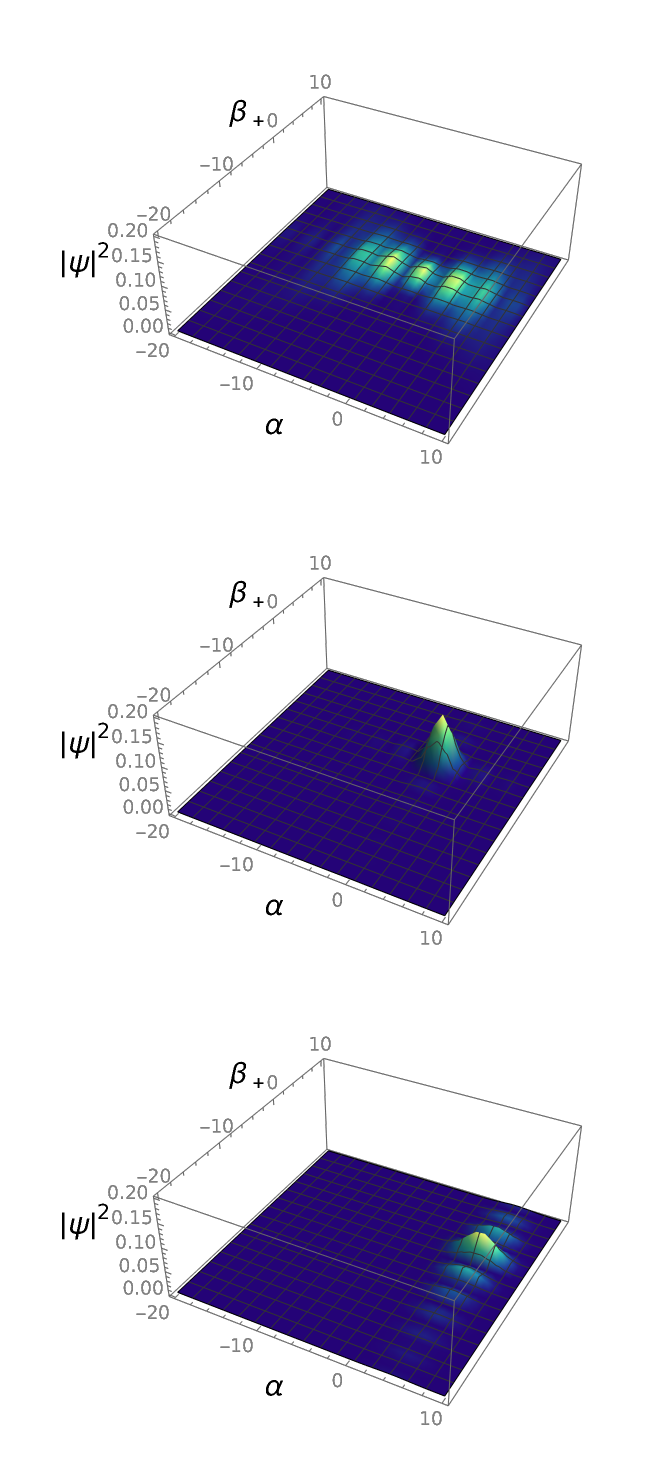} 
	\caption{Evolution of the wave packet $|\Psi|^2$ evaluated via numerical integration. Starting from the top, the probability density has been calculated for different values of the relational time $\xi=10,1,10^{-2}$ respectively, considering $\beta_-= 5$ and $\Lambda=10^{-18}$. An integration step equal to 0.1 has been used.}
	\label{conlambda}
\end{figure}

We aim to investigate if the on-set of a
inflating evolution is already able
to significantly modify the analysis
of the Bianchi I model in the vacuum, introducing a stabilization phenomenon for the classicalization tendency we discussed in the previous section.
Since we are in geometrical units we
adopt here a value of $10^{-18}$ for the cosmological constant, which is the expected relative value concerning a Planckian order of magnitude.
Hence, we proceed with the numerical evolution of the
probability density built as previously done where we choose Gaussian weights to peak the wave packets (\ref{prob}). In Fig.(\ref{conlambda}) we can easily see that the spontaneous tendency to localize happens faster and for a smaller $\xi$ value than the previous case. In other words, from Fig.(\ref{BianchiI}) one can notice that the process of localization appears in at least three orders of magnitude in the relational time while in the presence of the cosmological constant, it is already visible in one order of magnitude only.  A classical cosmology is reached but in the later evolution the spreading process of the wave packet is still present. The introduction of the cosmological constant seems not to be the mechanism to make the model a classical one.
Even if the analysis has been developed with a perturbative treatment, one can infer that new \lq \lq physics" has to be taken into consideration, in the case related to the specific chosen $f(R)$ and Bianchi model.

\section{Isotropic Universe dynamics}\label{FRLWsec}
As discussed in \cite{DeaFigMont}, considering an isotropic Universe in the presence of a scalar field $\phi$ and a cosmological constant $\Lambda$, a singular behaviour of the classical dynamics is recovered for a smaller value of $\xi$. In other words, the quantity $\partial \alpha / \partial \xi$ approaches $-\infty$ in correspondence to
the singular instant of the value 
\begin{equation}
\xi=\frac{p_{\alpha}^2}{2\left(\frac{3}{k}p_{\phi}^2+\Lambda e^{3\alpha}\right)}.
\end{equation}
In this sense, they argued that all of the Universe's inflationary evolution is contained in such an asymptotic behaviour since an asymptote for the function $\alpha(\xi)$ arises. 

Our analysis relies on deepening this concept and clarifies how, considering exclusively the cosmological constant, the quantum isotropic model is subjected to a stable process of localization, approaching a classical expanding Universe. To do that, we solve the Hamiltonian constraint with respect to the conjugate momentum of the time-like variable $\xi$ obtaining
\begin{equation}
p_{\xi}=\frac{p_{\alpha}+\sqrt{p_{\alpha}^2-6\xi\Lambda e^{3\alpha}}}{\xi}=-h_{\xi}.
\end{equation}
Hence, promoting again $h_{\xi}$ to a quantum operator
we achieve the dynamical evolution through the equation
\begin{equation}
i\xi\partial_{\xi}y(\xi)=\left(-k_{\alpha}-\sqrt{k_{\alpha}^2-6\xi \Lambda e^{3\alpha}}\right)y(\xi).
\end{equation}

Now, focusing our attention on the onset of the inflationary phase in which the condition in Sec.(\ref{secIII}) can be still retained as valid and according to the previous analysis of the Bianchi I model, we expand the square root argument in the Scrh\"odinger equation so that it reduces to the form


\begin{equation}
i\xi \partial_{\xi}y(\xi)=\left(-2k_{\alpha}+\frac{3\xi\Lambda e^{3\alpha}}{k_{\alpha}}\right)y(\xi),
\end{equation}
whose solution is
\begin{equation}
y(\xi)=\mathcal{D} e^{i\left(-\frac{3\xi\Lambda e^{3\alpha}}{k_{\alpha}}+2ik_{\alpha}\log(\xi)\right)},
\end{equation}
in which $\mathcal{D}$ is an integration constant.
Therefore, the wave function we obtain to compute the probability density becomes

\begin{equation}
\Psi(\alpha,\xi)=\int^{+\infty}_{-\infty}dk_{\alpha}A(k_{\alpha})e^{ik_{\alpha}\alpha}y(\xi).
\end{equation}

\begin{figure} [h!]
	\centering
	\includegraphics[width=9cm]{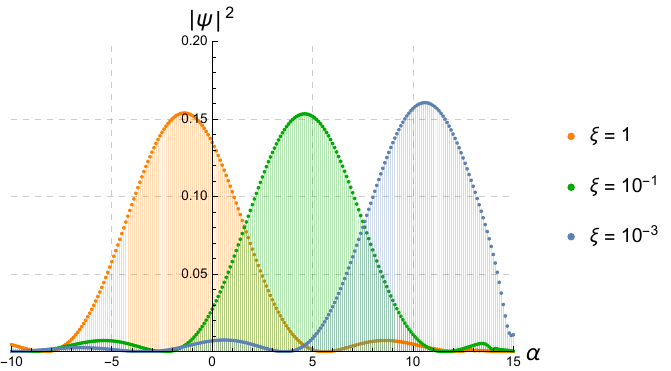} 
	\caption{Evolution of the wave packet $|\Psi|^2$ evaluated via numerical integration for different values of the relational time $\xi$, with an integration step equal to 0.1.}
	\label{frlw}
\end{figure}

Looking at Fig.(\ref{frlw}), it is important to note that starting with suitable initial conditions, \emph{i.e.} a well-shaped Gaussian distribution, the evolution of $|\Psi|^2$ approaches a stable peaked configuration, namely the classicalization.

It is worth stressing that when the cosmological constant is present, even if the values of $\alpha$ are noticeably increasing during the time evolution of the probability density and so $e^{3\alpha}$, condition \eqref{demo6} is retained valid for the whole phase.
Hence, a classical isotropic inflating Universe is recovered. 
This result validates the idea proposed in \cite{DeaFigMont} that the presence of a cosmological constant can stabilize the classicalization process of the isotropic Universe. Thus, we are naturally led to infer that also Bianchi I dynamics could be made definitely classical if the cosmological constant can sufficiently suppress the anisotropy degrees of freedom. However, the analysis of a full inflationary regime in the Bianchi I model could imply the necessity to take into account the explicit form of the potential $V(\xi)$, \emph{i.e.} of the considered modified gravity theory. This study is clearly of interest but out of the scope of the present analysis.


\section{Concluding remarks}\label{secVII}

The $f(R)$ theory of gravity, as viewed in the \textit{Jordan frame}, offers a natural framework to search for an internal clock in quantum dynamics, here identified with the non-minimally coupled scalar field to the tensor structure of the model. 

We generalize and complete the analysis originally developed in \cite{DeaFigMont} by considering the Bianchi Universe and, under suitable conditions (\emph{i.e.} neglecting the potential term of the scalar field and the spatial curvature of the model), we restrict the analysis to the quantum dynamics of a Bianchi I cosmology. 
We showed how both in the absence 
or in the presence of the cosmological 
constant (here treated as a perturbative phenomenon to the quantum dynamics), 
the behavior of a wave packet, solution of the Schr\"odinger equation, is 
characterized by a spontaneous tendency to 
localize, \emph{i.e.} a classical cosmology 
is approached. However, this quasi-classical state is not dynamically stable, since in the later evolution the wave packets are subjected to a spreading process.  For the same behavior observed in the case of an isotropic Universe, in the presence of a scalar field in \cite{DeaFigMont} it was argued that the inflationary scenario could provide the 
stabilization mechanism of the classical 
behavior, occurring at a given value of 
the relational time. 
The analysis here developed in Sec.(\ref{FRLWsec}) 
demonstrated the validity of this guess implying that 
a classical isotropic inflating Universe emerges considering a perturbative cosmological constant.
The same result is not proven by our 
study of the Bianchi I model, since 
the presence of the cosmological constant seems unable to provide a stable 
localization of the quantum Universe. 
Actually, we can describe only 
the onset phase of the inflation, since the analysis of the wave packets 
has been numerically available only 
when the cosmological constant term is 
a weak perturbation of the free dynamics and only in this limit we can be sure that the spatial curvature $V_B(\beta_+,\beta_-)$ is negligible.
We can infer that a more complete study could clarify how the inflation 
dynamics is able to induce a decay of 
the quantum anisotropy, marked by the 
behavior of their mean values and 
standard deviations. If such a process 
was able to induce an isotropic-like 
Universe, then the stabilization of the 
classical dynamics would be reached according to the same picture of Sec.(\ref{FRLWsec}), as 
discussed above.

The relevance of the scenario we discussed in the present study relies on the possibility to extend this dynamical scheme and all its implication in the 
context of a generic cosmological solution \cite{primordial} by implementing on a quantum level the so-called \lq\lq BKL-conjecture" \cite{BKL82,montani95}. The idea emerging from our study is that in a $f(R)$ under certain conditions, a generic inhomogeneous Universe 
could spontaneously evolve in the natural internal time $\xi$ from a quantum behavior to a classical quasi-isotropic inflating regime \cite{ImponenteMontani2001,KL63}.


%

\end{document}